\documentclass{aa}  
\usepackage{graphicx}
\usepackage{txfonts}
\usepackage{orcidlink}
\usepackage{textcomp}

\newcommand{\speed}[1]{#1 km~s${}^{-1}$}

\usepackage{txfonts}

\begin{document} 

\title{Resolving interchange reconnection dynamics in a fan-spine-like topology observed by Solar Orbiter}

\author{
          Yadan Duan~\orcidlink{0000-0001-9491-699X}
          \inst{1,2,3}\thanks{Corresponding author; duanyadan@ynao.ac.cn}
           \and
           Xiaoli Yan~\orcidlink{0000-0003-2891-6267}\inst{1,2}
           \and
          Junchao Hong~\orcidlink{0000-0002-3804-7395}\inst{1}
          \and
          Hechao Chen~\orcidlink{0000-0001-7866-4358}\inst{4,2}
          \and
          Yuhang Gao~\orcidlink{0000-0002-6641-8034}\inst{3,5}
          \and
          Zheng Sun~\orcidlink{0000-0001-5657-7587}\inst{3,6}
          \and
          Zhenyong, Hou~\orcidlink{0000-0003-4804-5673}\inst{3}
          \and
          Jincheng Wang~\orcidlink{0000-0003-4393-9731}\inst{1,2}
            }

   \institute{Yunnan Observatories, Chinese Academy of Sciences, Kunming, 650216, China\\
                 \email{duanyadan@ynao.ac.cn}
         \and
            Yunnan Key Laboratory of Solar Physics and Space Science, Kunming, 650216, People's Republic of China\
         \and
          School of Earth and Space Sciences, Peking University, Beijing, 100871, People's Republic of China\
         \and
          School of Physics and Astronomy, Yunnan University, Kunming, 650500,  China\
         \and
         Centre for Mathematical Plasma Astrophysics, Department of Mathematics, KU Leuven, Celestijnenlaan 200B bus    2400, B-3001 Leuven, Belgium\
         \and
         Leibniz Institute for Astrophysics Potsdam, An der Sternwarte 16, Potsdam 14482, Germany\
         }

   \date{Submit}

  \abstract
{Interchange reconnection is thought to play a significant role in the production of solar jets and the solar wind. The dynamics of interchange reconnection in the low corona might be more complex than recognized before in higher temporal and spatial resolutions, however. Using unprecedentedly high-resolution observations from the Extreme Ultraviolet Imager (EUI) on board the Solar Orbiter, we analyzed the dynamics of interchange reconnection in a small-scale fan-spine-like topology. The interchange reconnection that continuously occurs around the multinull points of the fan-spine-like system exhibits a quasi-periodicity of $\sim$ 200 s that nearly covers the entire evolution of this system. Continuous evolution and a reversal of multiple current sheets are observed over time near the null point. These results reveal that the dynamics of interchange reconnection is likely modulated by emerging magnetic structures, such as mini-filaments and emerging arcades. Moreover, a curtain-like feature with a width of 1.7 Mm is also observed near the interchange reconnection region and persistently generates outflows. This is similar to the separatrix curtain reported in the pseudostreamer structure. This study not only demonstrates the complex and variable reconnection dynamics of interchange reconnection within a small-scale fan-spine topology, but also provides insights into the self-similarity of magnetic field configurations across multiple temporal and spatial scales.}

\keywords{Sun: activity -- Sun: corona -- Sun: magnetic topology -- magnetic reconnection}

   \maketitle

\section{Introduction}

Interchange reconnection represents a fundamental physical process in the solar atmosphere. It is characterized by the reconnection between closed magnetic field lines and adjacent open magnetic fields of opposite polarity \citep{1998SSRv...86...51F,2002JGRA..107.1028C}. It can lead to a rapid conversion of magnetic energy that enables the transfer of mass and energy along magnetic field lines. Meanwhile, interchange reconnection can cause plasma heating and particle acceleration, which allows the escape of energetic electrons into interplanetary space along open magnetic field lines. Additionally, interchange reconnection is also an effective way to release the magnetic twist from closed magnetic fields into interplanetary space \citep[e.g.,][]{2019Natur.576..237B,2020ApJ...896L..18S,2020JPhCS1620a2020S}. In the solar corona, diverse magnetic field configurations are conducive to the occurrence of interchange reconnection. It is therefore crucial to characterize the interchange reconnection dynamics in different magnetic field configurations for understanding the triggering mechanisms for solar jets \citep[e.g.,][]{1992PASJ...44L.173S,2011ApJ...742...82K} and the origin and properties of the solar wind \citep[e.g.,][]{2014ApJ...787..145M,2023A&A...675A..55P}.

\par 
Pseudostreamers are large-scale coronal structures and provide a favorable environment for the occurrence of interchange reconnection. Pseudostreamers separate open field lines of the same polarity and contain two closed magnetic fluxes below their cusp \citep{2007ApJ...658.1340W} with a $\pitchfork$-type structure. The pseudostreamers are usually located more than 1.5 solar radii above the solar surface and are formed by multiple null points linked by separators \citep[e.g.,][]{2011ApJ...731..111T,2012ApJ...759...70T,2013ApJ...777...72S,2014ApJ...787..145M}. In this configuration, the gradient of magnetic connectivity among open and closed magnetic fields is strong \citep[e.g.,][]{2014ApJ...787..145M,2021ApJ...913...64S,2023A&A...675A..55P}. This is known as a separatrix-web (S-Web) and is predicted to favor the occurrence of interchange reconnection. The S-Web model was proposed by \citet{2011ApJ...731..112A}, \citet{2011ApJ...731..111T}, and \citet{2011ApJ...731..110L}, who showed a network of separatrix surfaces and quasi-separatrix layers (QSLs). Simulation studies demonstrated that a vertical fan surface emanates from the null points. This is referred to as a separatrix curtain because it has a curtain-like shape \citep[e.g.,][]{2014ApJ...787..145M,2021ApJ...913...64S,2022LRSP...19....1P}. The separatrix curtain is surrounded by QSLs, which results from the rapid divergence of the field lines in the vicinity of the null points associated with the separatrix surfaces \citep{2011ApJ...731..111T}. The interchange reconnection that occurs in pseudostreamers is a gradual physical process \citep{2014ApJ...787..145M} that can last for several hours without any drastic changes \citep[e.g.,][]{2013ApJ...777...72S}. The pseudostreamer topology, which probably more accurately reflects the actual magnetic structure of the system, is thought to play a significant role in the generation of the slow solar wind. It is therefore crucial to understand the dynamics of this system for the analysis of the mass and energy release of the slow solar wind. Recently, \citet{2023NatAs...7..133C} used middle coronal observations to investigate the dynamics of pseudostreamer systems, providing observational evidence that the reconnection dynamics along the S-Web are responsible for the release of slow solar wind plasma. However, it remains challenging to observe this dynamic process of the pseudostreamer system in the high corona.

\par
The fan-spine structure is defined by a dome-shaped fan separatrix surface, a magnetic null point positioned at its apex, and the inner and outer spines that pass through the null point \citep[e.g.,][]{2002A&ARv..10..313P,2021RSPSA.47700949L,2022LRSP...19....1P}. Similar to the large-scale pseudostreamers, a smaller-scale fan-spine structure is also a favorable place for the occurrence of interchange reconnection.
Recent observations revealed that the magnetic null-point position of fan-spine configurations is approximately 1$-$6 Mm above the solar surface \citep[e.g.,][]{2023NatCo..14.2107C,2024ApJ...962L..38D}.
The fan-spine topology is thus associated with various small-scale solar phenomena, such as solar jets \citep[e.g.,][]{2011ApJ...728..103L,2012ApJ...760..101W,2017ApJ...835...35H,2019ApJ...885L..11S,2022ApJ...926L..39D}, solar flares \citep[e.g.,][]{2013ApJ...778..139S,2018ApJ...859..122L,2018ApJ...867..159S,2019ApJ...871..105Z,2020ApJ...898..101Y}, ultraviolet bursts \citep[e.g.,][]{2017A&A...605A..49C,2019ScChE..62.1555C,2025A&A...693A.221B}, Ellerman bombs \citep[e.g.,][]{2025A&A...698A.174B}, macrospicules \citep[e.g.,][]{2023ApJ...942L..22D}, small-scale spiral jets \citep[e.g.,][]{2023NatCo..14.2107C,2025A&A...696L...2L}, jetlets \citep[e.g.,][]{2022ApJ...933...21K,2025ApJ...987..193Y},  and coronal bright points \citep[e.g.,][]{2012ApJ...746...19Z,2022ApJ...935L..21N,2024A&A...687A.171F}. An increasing number of observations and simulations currently support that jets and flares can result from magnetic reconnection between the magnetic flux rope (or filament) that is initially confined below the fan and the magnetic field of the outer fan \citep[e.g.,][]{2013ApJ...771L..30J,2018ApJ...854...64S,2017Natur.544..452W,2018ApJ...852...98W,2020ApJ...900..158Y,2021ApJ...923...45Z,2024ApJ...962L..38D,2024ApJ...976..135Y}. 
After the eruption, the remains of an inner bright patch encircled by a circular flare ribbon are observed on the solar surface \citep[e.g.,][]{2009ApJ...700..559M,2015ApJ...812L..19L,2019ApJ...871....4H,2022SoPh..297....2N,2024RvMPP...8....7Z}, and these features indicate the footpoint positions of the inner spine and the dome-like fan separatrix surface. The lifetime of circular ribbon flares ranges from 4 to 205 minutes, with a mean value of approximately 50 minutes \citep{2022ApJS..260...19Z}. The speeds and ejection heights of solar jets can reach several hundred kilometers per second and several hundred megameters, respectively.

\par
Interchange reconnection around the null point in the fan-spine system is also an active field of theoretical and numerical research. It helps us to understand the transfer of magnetic flux across different topological regions \citep[e.g.,][]{2009ApJ...691...61P,2012SoPh..276..199M,2013A&A...554A.145F,2013ApJ...774..154P,2022LRSP...19....1P}. Traditional studies have generally considered that fan-spine structures contain a single null point \citep[e.g.,][]{2009ApJ...691...61P,2017Natur.544..452W}, which differs from pseudostreamer structures, which are formed by several null points linked by separators \citep[e.g.,][]{2012ApJ...759...70T}. However, \citet{2024ApJ...962L..38D} recently reported that the fan-spine topology probably contains more than one null point (potentially multiple null points, QSLs, or separators). \citet{2024ApJ...962L..38D} reported that the position of a current sheet is along the QSL that connects two null points. Moreover, small-scale evolving jets also exhibit various dynamic processes at their base, such as multiple small-scale eruptions, plasma flows, and moving blobs \citep[e.g.,][]{2022A&A...664A..28M,2023ApJ...944...19L}. Thus, the dynamic of interchange reconnection within small-scale fan-spine structures is probably more complex and requires further investigations with higher resolution. 

\par
We used data with a high temporal and spatial resolution acquired by the Extreme Ultraviolet Imager (EUI; \citealt{2020A&A...642A...8R}) on board Solar Orbiter (\citealt{2020A&A...642A...1M}) to investigate the dynamics of interchange reconnection that continuously occurs around the null point of a fan-spine-like structure. Moreover, we also identify a curtain-like feature similar to the large-scale pseudostreamer within a small-scale fan-spine-like structure. 

\section{Observations}
On 2024 April 5, Solar Orbiter was positioned approximately 0.29 AU from the Sun at an angle of 83 degrees between Solar Orbiter and Earth relative to the Sun \citep[e.g.,][]{2025arXiv250904771T}. The event of interest occurred at the western limb of the Sun as observed by Solar Orbiter, rendering it completely invisible from Earth. Consequently, all data we used were obtained from Solar Orbiter. The EUI on board Solar Orbiter consists of three telescopes: the Full Sun Imager (FSI), and two High-Resolution Imagers (HRI\textsubscript{EUV} and HRI\textsubscript{Ly$\alpha$}). HRI\textsubscript{EUV} 174 \AA~ images provide the dynamic evolution of the fan-spine-like structure and the complex interchange reconnection process around the null point within the fan-spine-like system. The time cadence of HRI\textsubscript{EUV} 174 \AA~ observations is 16 s, and the pixel scale is 0.\arcsec 49, with an effective exposure time of 2 s. A pixel corresponds to 0.108 × 0.108 Mm$^2$ on the Sun. The imaging observations in two channels (FSI 174 \AA~and FSI 304 \AA~) from the FSI also enable us to study the plasma motions from a larger field of view. The temporal cadence and spatial pixel scale of FSI data are 10 minutes and 4.44\arcsec. We used a cross-correlation for the data alignment and jitter elimination.

\section{Results}
\begin{figure*}
\centering
\includegraphics[scale=0.5]{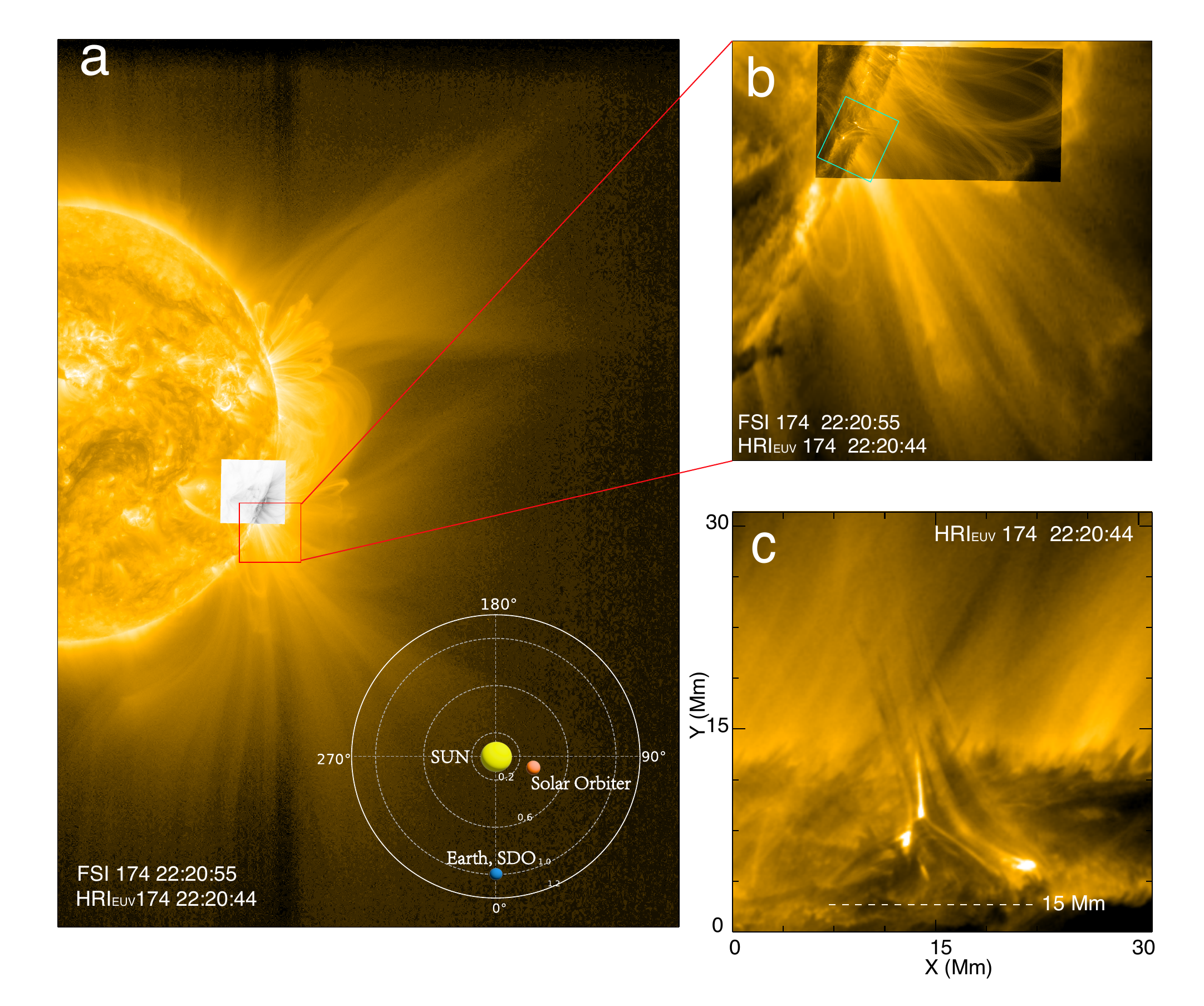}
\caption{\textbf{a}: Positions of Solar Orbiter and Earth at 22:20 UT on 2024 April 5. The HRI\textsubscript{EUV} inverted grayscale is overlaid on the FSI$_{174}$ image. It shows the location of a fan-spine-like structure at this time. The field of view in panel (b) corresponds to the red box indicated in panel (a), showing the HRI overlaid on the FSI image at 22:20 UT \textbf{c}: Zoom-in of HRI\textsubscript{EUV} 174 \AA~( cyan box in \textbf{b}) showing the structure at the solar limb at a scale of about 15 Mm. 
\label{fig:general}}
\end{figure*}

Figures 1(a) and 1(b) present an overview of a fan-spine-like structure observed at the solar limb from the perspective of Solar Orbiter. A closer view of this structure, indicated by the cyan box in Fig. 1(b), is shown in Fig. 1(c) after a 115° counterclockwise rotation to achieve an upright orientation. Figure 1(c) reveals that the scale of the fan dome is approximately 15 Mm at the solar limb. At this time, the morphology of the plasma outflows along the outer spine of the fan-spine system is that of an anemone solar jet, characterized by an inverted-Y shape comprising a bright, straight jet spire and a jet base. This inverted-Y feature is thought to be the consequence of reconnection between small-scale magnetic arcades and the unipolar background fields \citep[e.g.,][]{2014Sci...346A.315T,2018ApJ...854...92T}. Given the location of this structure at the solar limb and the associated projection effects, we did not use PHI \citep{2020A&A...642A..11S} to extrapolate the magnetic field \citep[e.g.,][]{2024A&A...687A..13P} and obtain the fan-spine feature. In the context of magnetic reconnection in three-dimensional geometries, the simplest anticipated outcome of flux emergence into a locally unipolar region is the formation of a fan-spine magnetic field configuration \citep[e.g.,][]{2009ApJ...704..485T}. Thus, the fan-spine topology represents the 3D magnetic structure of the straight anemone jet \citep[e.g.,][]{2021RSPSA.47700217S}.

\begin{figure*}
\centering
\includegraphics[scale=0.25]{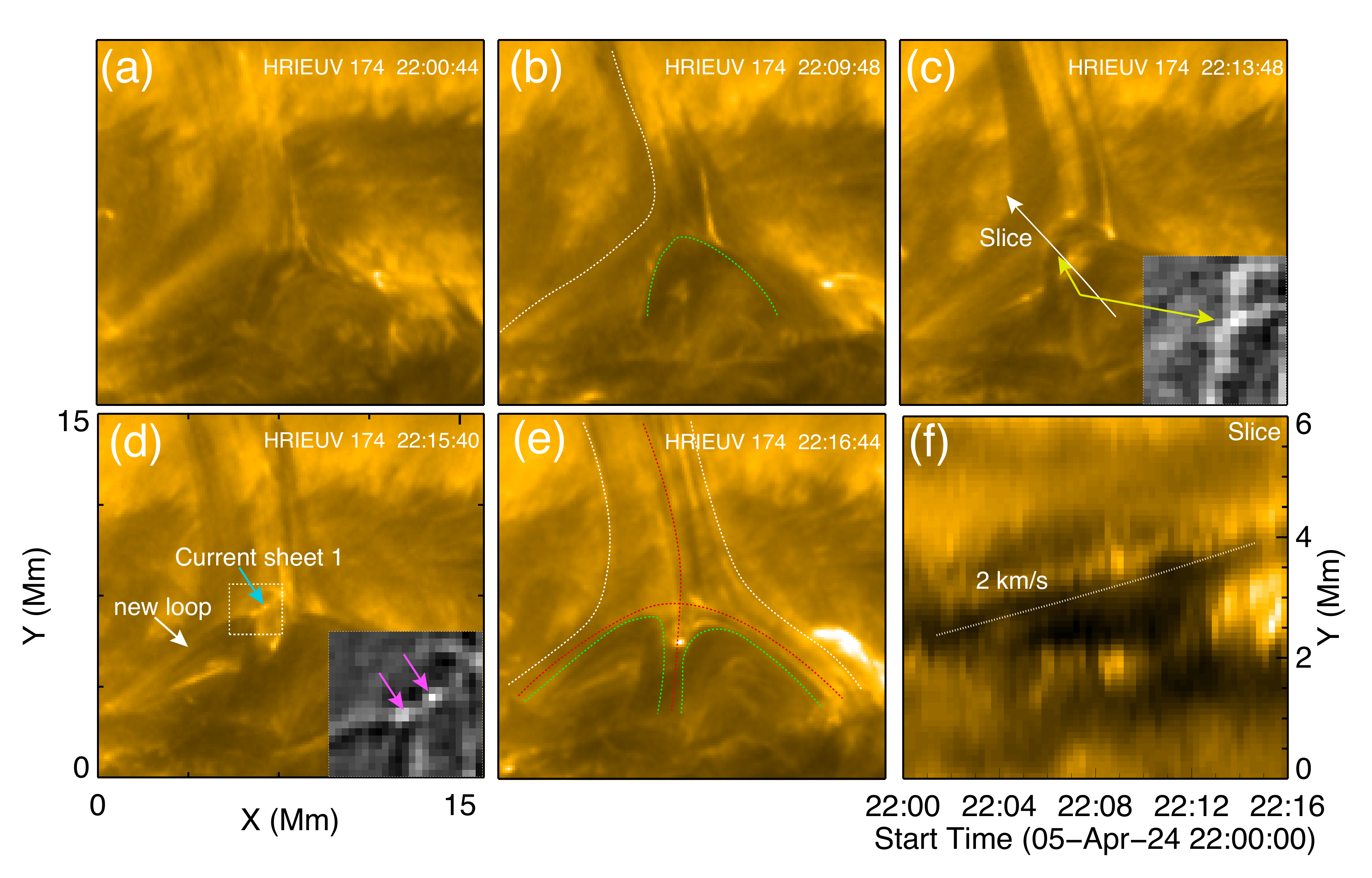}
\caption{(a)$-$(e): HRI\textsubscript{EUV} 174\AA~sequence images depicting the evolution process of the fan-spine-like topology. In panel (b), the emerged and ambient open field are outlined by dashed green and white lines, respectively. The yellow arrows point to the X point of reconnection between the emerging field and the surrounding open magnetic field in panel (c). The pink arrows denote the plasma blobs in current sheet 1 (d). Panel (e): Fan surface and spine of the fan-spine-like structure are outlined by dashed red lines, and the dashed green and white lines outline the closed magnetic flux under the fan dome and the open flux above the fan dome, respectively. Panel (f): Time-distance diagram along the direction of the white line in panel (c). An animation of the whole evolution process of the fan-spine system is available from 22:00 UT to 22:58 UT. The duration of the animation is 9 s.
\label{fig:general}}
\end{figure*}

Figure 2 shows the evolution of the fan-spine-like topology as observed in HRI\textsubscript{EUV} 174\AA~ images. At approximately 22:00:44 UT, a small dark closed-loop system emerges to the right of the coronal plume region. This dark loop might represent an arch filament system carrying cool material \citep[e.g.,][]{1967SoPh....2..451B,2018SoPh..293...93C,2018ApJ...855...77S}. By 22:09:48 UT, the emerged loop gradually expanded toward the background coronal loop region. The speed at which the magnetic loop carrying cold material approached the background field is approximately 2 $\speed$, as shown in Fig. 2(f). At 22:13:48 UT, an X-point, the signal of interaction reconnection between two systems at the reconnection site, is marked by the yellow arrows. This interaction results in the formation of a new closed loop (denoted by the white arrow), as shown in Fig. 2(d). Afterward, a weak sheet-like feature developed near the X-point and connected the new loop and the open field, as denoted by the cyan arrow in Fig. 2(d). This sheet-like structure delineates the current sheet associated with the reconnection configuration, as suggested by previous observational studies  \citep[e.g.,][]{2014ApJ...797L..14T,2016NatCo...711837X,2017ApJ...851...67S,2019ApJ...879...74C,2021ApJ...915...39H,2022ApJ...924L...7L,2024ApJ...974..104D,2024ApJ...962L..38D}. Moreover, bright plasma blobs are also observed within the current sheet, as indicated by the pink arrows in Fig. 2(d) and reported by previous studies \citep[e.g,][]{2016NatPh..12..847L,2022NatCo..13..640Y,2024A&A...687A.190H,2025A&A...702A.188N,2025ApJ...990L..16T}. The plasma blobs might be generated during the interchange reconnection as a result of a tearing instability, as indicated by the simulation studies conducted by \citet{2009PhPl...16k2102B}. As the interaction becomes more dynamic, a common footpoint region shared by the emerged loop and the newly formed loops is observed that collectively displays a fan-spine-like morphology, as outlined by the dashed lines in Fig. 2(e). The interaction between this newly emerging magnetic flux and the adjacent background field evolves into a fan-spine morphology, consistent with \citet{2009ApJ...704..485T} and \citet{2024ApJ...962L..38D}. Because of the position of the event at the solar western limb, which lacks magnetic field data, we cannot extrapolate the magnetic field for this event. Thus, we refer to this event as the fan-spine-like structure.
\par
Figure 3 presents the observational dynamic features associated with the transition from current sheet 2 (CS2) to current sheet 3 (CS3). At about 22:17:16, a bright emission appears beneath filament 1, located at the side of the fan-spine-like structure, and the CS2 feature appears within the fan-spine-like structure. Although the EUI image only provides the projected trajectory of filament 1 and CS2, they clearly moved upward together (see the animation). The upward movement of the CS2 might be a reflection of larger-scale magnetic field relaxation in this region. Ultimately, filament 1 erupted, as evidenced by its structural changes from a closed magnetic flux system to an open field, shown in Fig. 3(c). Another filament (filament 2) is observed at 22:21:32 UT. About two minutes later, filament 2 was activated and erupted in the direction of CS2. This may be triggered by CS2 and effectively eliminates the restraining magnetic arcade overlaying the underlying filament 2. At 22:24:12 UT, a bright linear feature, indicated by the white arrow, connected the base of the jet spire and the top of the post-flare arcade. This feature is similar to the flare current sheet that formed in the wake of the filament eruption\citep{1999ApJ...510..485A,2017Natur.544..452W} and is referred to as CS3. Figures 3(d1)$-$3(f1) show the schematic diagrams corresponding to panels (d)$-$(f) and the possible connectivity of filament 2.  
\par

In the time–distance diagram along the arrow S3 (see Fig. 3(f)), plasma outflows occur continuously and a significant jet erupts ($\sim$ 22:23 UT), as displayed in Fig. 3(i3). We also employed the wavelet and Fourier analysis method \citep{1998BAMS...79...61T,2016ApJ...825..110A} to obtain the periods of the continuous plasma outflows. We extracted a light curve from S3 (the location is marked by the dashed red line in Fig. 3(i3)). The result is illustrated in Figs. 7(a1)$-$7(a3). The speeds of these plasma outflows are presented in the time-distance diagram along the ejection to the highest position, as shown in Figs. 6 (a) and 6(d).

\begin{figure*}
\centering
\includegraphics[scale=0.7]{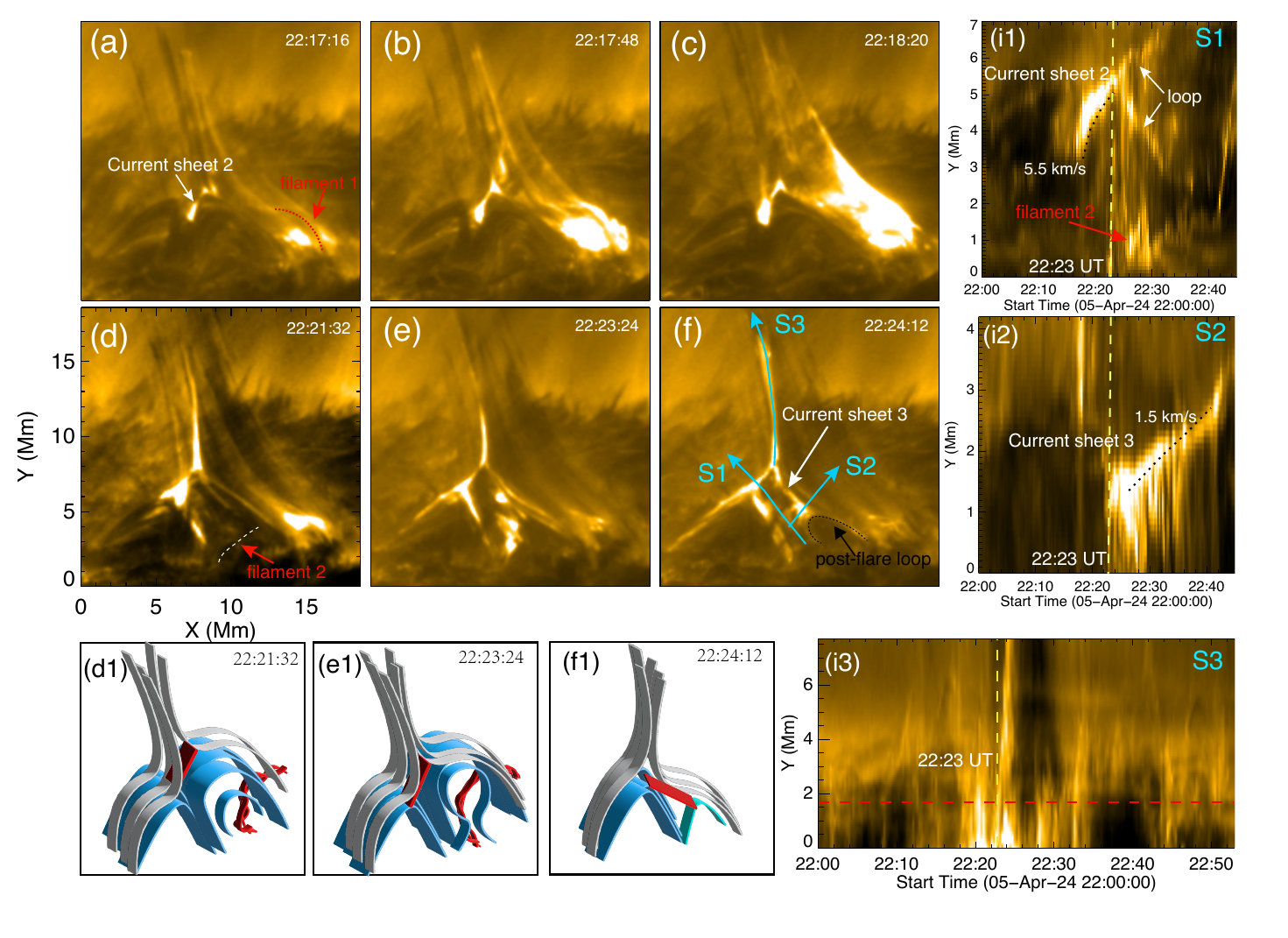}
\caption{(a)$-$(f): HRI\textsubscript{EUV} 174\AA~sequence images showing the production of two current sheets. (d1)$-$(f1): Schematic diagrams of the evolutionary sequence of filament 2, corresponding to the respective time instances shown in panels (d)$-$(f). 
(i1)$-$(i3): Time-distance diagrams along the direction of the cyan arrows in panel (f). The dashed black line outlines the post-flare loop after the eruption of filament 2 in panel (f). The dashed red line in panel (i3) represents the position from which the light curve was extracted for wavelet analysis, with the results presented in Figs. 7(a1)$-$7(a3).
\label{fig:general}}
\end{figure*}
\par
Figures 3(i1) and 3(i2) show the time–distance diagrams along the cyan arrows S1 and S2 in panel (f). At around 22:23 UT, CS2 clearly transformed into a two-loop system moving in different directions (see the animation), while CS3 appeared. At this moment, filament 2 erupted toward CS2, as indicated by the red arrow in Fig. 3(i1). CS2 was observed from approximately 22:17 UT to 22:23 UT, during which its width ranged from 0.6 to 0.8 Mm, which is consistent with prior statistical analyses \citep{2024ApJ...974..104D,2024ApJ...964...58D}. The length of CS2 appeared to grow over time from an initial 3 Mm to a final 5 Mm before it disappeared. According to the Sweet-Parker mechanism \citep{2000mare.book.....P}, the reconnection rate (the inflow Alfv\'en Mach number (M$_A$)) is theoretically proportional to the ratio of the current sheet thickness (d) to its length \citep[L; e.g.,][]{2016NatCo...711837X,2019ApJ...879...74C,2024MNRAS.528.1094Y}. Based on these measurements, the reconnection rate of the CS2 was estimated to be in the range of 0.12 to 0.2. During the period from 22:23 to 22:40 UT, the width of CS3 was approximately 0.4 Mm to 0.6 Mm, and its length varied from 2.7 Mm to 7.4 Mm. Unlike CS2, CS3 contracted for a time following its initial elongation. The estimated reconnection rate of CS3 ranged from 0.12 to 0.17. Both current sheets showed an upward movement trend, and CS2 and CS3 moved at speeds of 5.5 km/s and 1.5 km/s, respectively.  

\begin{figure*}
\centering
\includegraphics[scale=0.4]{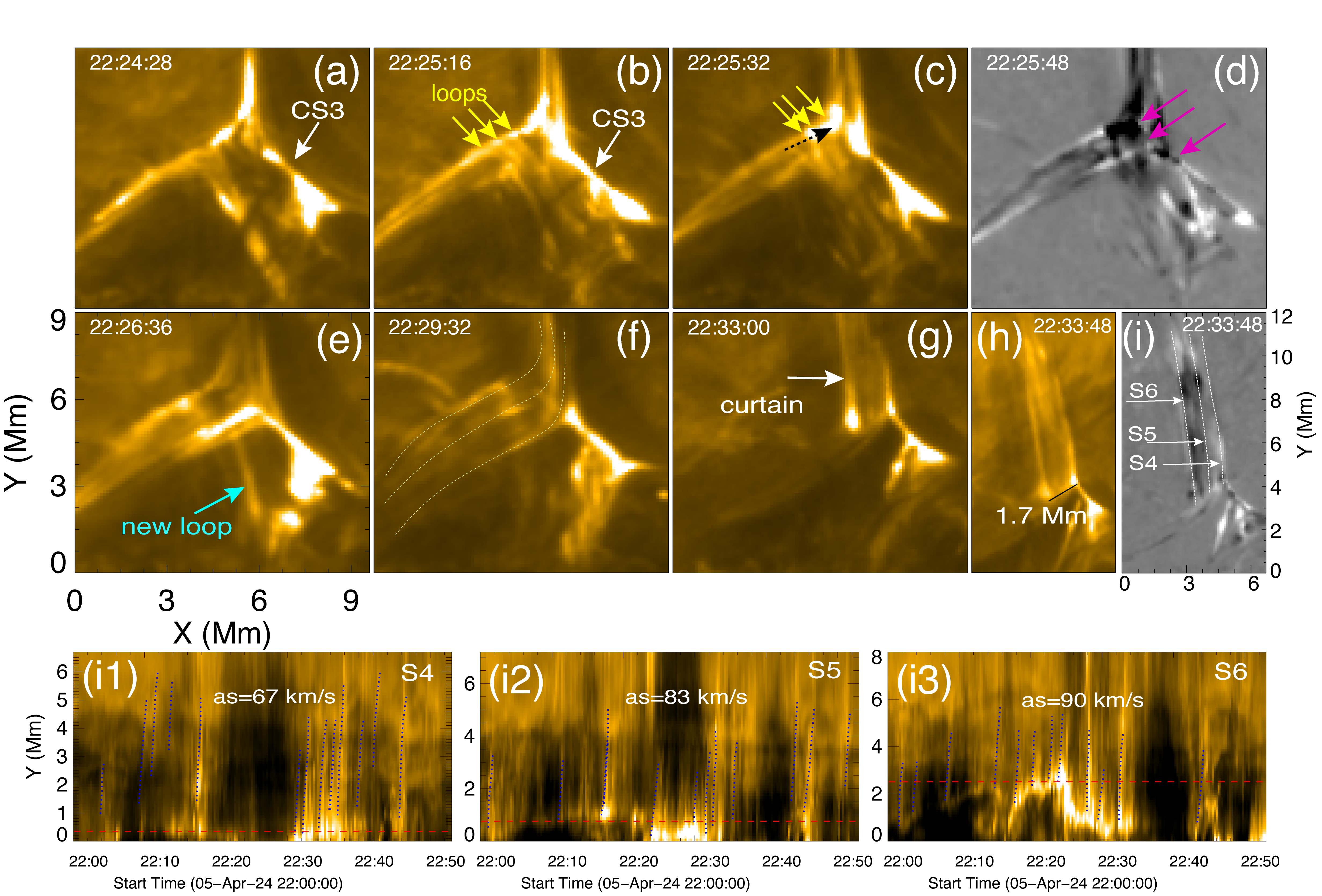}
\caption{(a)$-$(f): Sequence of HRI\textsubscript{EUV} 174 Å images illustrating the continuous interchange reconnection process. (b) Three groups of loops (yellow arrows). (c): The dashed black arrow shows the slipping motion of these three loops. (d): Reconnection site between these three loops and the ambient open fields (pink arrows). Open magnetic field lines of the three loops after the reconnection in panel (f) (dashed yellow lines). Panels (g)$-$(i) show a curtain-like feature. (i1)$-$(i3): Time-distance diagrams along the direction of the dashed white lines in panel \textbf{(i)}. The dotted blue lines in panels (i1)$-$(i3) trace the inclined ridges that correspond to the plasma outflows along the curtain feature. The speeds were calculated by performing a linear fit on the height-time measurements. The “as” is an abbreviation for “average speed”. The dashed red lines in panels (i1)$-$(i3) mark the locations from which the light curves were extracted for wavelet analysis, with the results presented in Figs. 7 (b) to 7(d).
\label{fig:general}}
\end{figure*}
\par
Figure 4 illustrates the dynamic evolution of the interchange reconnection in the outflow region of CS3. A set of magnetic loops is clearly visible to the left of the jet spire, as shown in Fig. 4(a). These loops are reconnected loops that formed in the outflow region of CS2. At 22:25:16 UT, this set of magnetic loops clearly showed three distinct groups of field lines (indicated by the yellow arrows). Between 22:24:28 UT and 22:25:32 UT, these loops moved toward the jet spire in a slipping motion. They reconnected with the surrounding open field lines at CS3 (the reconnection point is indicated by the purple arrow in panel (d)). At 22:26:36, the newly formed loops after interchange reconnection are shown by the cyan arrow. The closed field lines on the left of the jet spire eventually changed into open field lines through the dynamic evolution of interchange reconnection in a curtain-like feature (see panels (f)$-$(i)). The width of this curtain-like feature is approximately 1.7 Mm. The time-distance diagram along the curtain-like structure from arrows S4$-$S6 shows the continuous plasma outflows along the curtain feature. The average velocity of these plasma outflows is $\sim$ 80 $\speed$. We extracted the light curves from S4, S5, and S6 at the location marked by the dashed red lines in Figs. 4(i1)$-4$(i3). The results of the wavelet and Fourier analysis are shown in Figs. 7(b)$-$7(d).

\begin{figure*}
\centering
\includegraphics[scale=0.9]{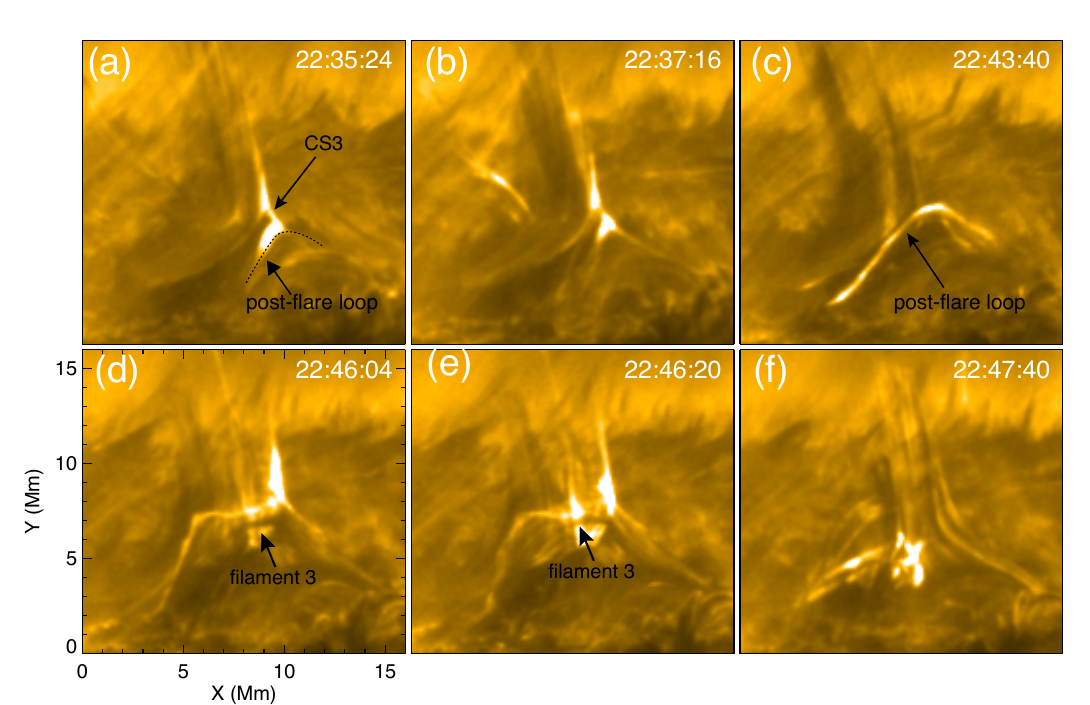}
\caption{Panels (a)$-$(f): Final evolution process of the fan-spine-like system. The black arrows denote another filament (filament 3) in panels (d)$-$(e).
\label{fig:general}}
\end{figure*}
\par
Figure 5 shows the ultimate breakdown process of this fan-spine-like system. CS3 gradually approached the post-flare loop, as shown in Figs. 5(a)$-$5(c). This might be caused by the migration of the null point in the fan-spine system as a result of changes in the gas pressure and magnetic pressure. Subsequently, another filament, referred to as filament 3, appeared below the post-flare loop, as displayed in Fig. 5(d). Together, they showed dynamic evolutionary characteristics of uplift and interacted with each other (see Fig. 5(e)). The post-flare loop and filament 3 eventually transitioned from a closed magnetic flux to an open flux at 22:47:40 UT.

\begin{figure*}
\centering
\includegraphics[scale=0.55]{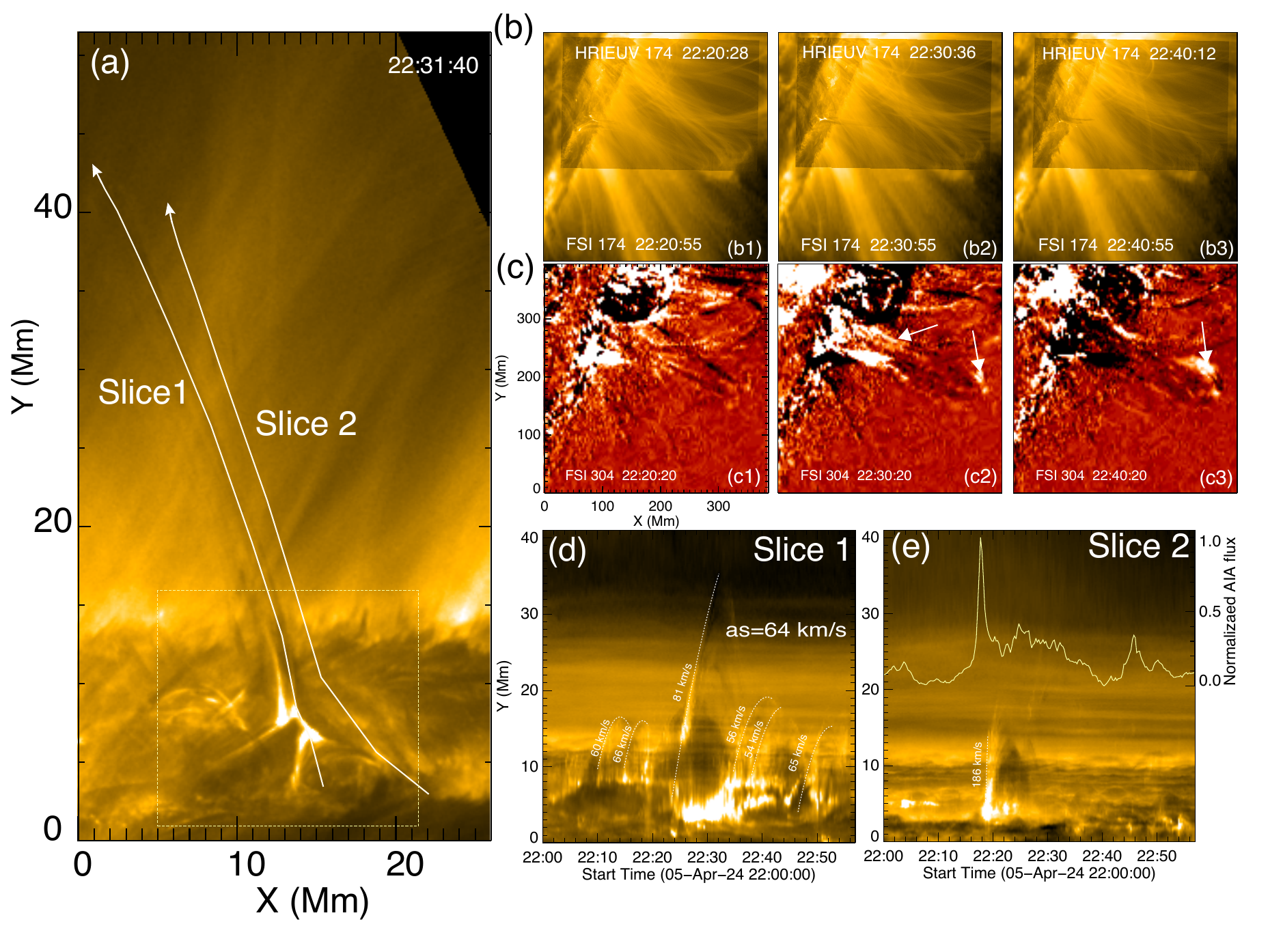}
\caption{Panel (a): Direction in which plasma outflow is ejected to its highest position (white arrow, slice 1). Slice 2 is along the direction of the eruption of filament 1 at the side of the fan-spine structure.
Panels (b)$-$(c): HRI\textsubscript{EUV} images overlaid on the FSI$_{174}$ images and FSI$_{304}$ running-difference images, showing the location of the fan-spine-like structure. The white arrows illustrate the plasma ejection material propagating along the closed magnetic field lines in panels (c2)$-$(c3). Panels (d)$-$(e): Time$-$distance intensity diagrams of HRI\textsubscript{EUV} 174 \AA~plotted along slices 1 and 2 in panel (a). The yellow curve represents the EUI 174 \AA~light curve, which is calculated within the dashed yellow box in panel (a) from 22:00 UT to 22:57 UT. 
\label{fig:general}}
\end{figure*}

\par
Figure 6 illustrates the spatial relation between the fan-spine-like structure and the surrounding magnetic field. For clarity, we superposed HRI\textsubscript{EUV} 174 \AA~images on a sequence of FSI 174 \AA~images (see Figs. 6(b1)$-$6(b3)). Additionally, the running-difference images from the FSI 304 \AA~sequence, shown at 10-minute intervals, show the process of plasma outflow. To quantify the plasma motion, two cuts were placed, as shown in Fig. 6(a). Slice 1 was positioned along the path where the plasma ejection reached its maximum height within the fan-spine-like structure, and slice 2 was established in the direction of the eruption of filament 1 along the side of this structure. By calculating the light curve within the dashed yellow box in Fig. 6(a), we found that the peak of this intensity curve corresponds to the eruption of filament 1. Based on the time-distance diagram of Fig. 6(e), we determined that the time and speed of the erupted filament 1 are about 22:18 UT and 186 $\speed$. The average speed of the recurring plasma ejections is 64 $\speed$. The highest plasma ejection occurred around 22:23 UT (see Fig. 6(d)), corresponding to the eruption of filament 2 in Figs. 3(d)$-$3(f) and (i3). The FSI 304 \AA~running-difference images show that the plasma material flows along the closed field lines and finally impacts the solar surface, as denoted by white arrows in Figs. 6(c2) and 6(c3). No disturbance signals were observed in the HRI\textsubscript{EUV} 174 \AA~or FSI 174 \AA~images, indicating that the temperature of the material flows was not high relative to the coronal loop. The low temporal resolution of the FSI prevents us from distinguishing which filament eruption the material that propagated along the closed magnetic field lines in Figs. (c2)$-$(c3) originated. These results indicate, however, that the reconnection in the low corona can supply material and energy to the high corona. Additionally, the ejected plasma flows subsequently fall down, which might correspond to coronal rain within the coronal loops. The ongoing coronal rain indicates the spatial distribution of the coronal-loop heating \citep{2010ApJ...716..154A,2022FrASS...920116A}.

\begin{figure*}
\centering
\includegraphics[scale=0.58]{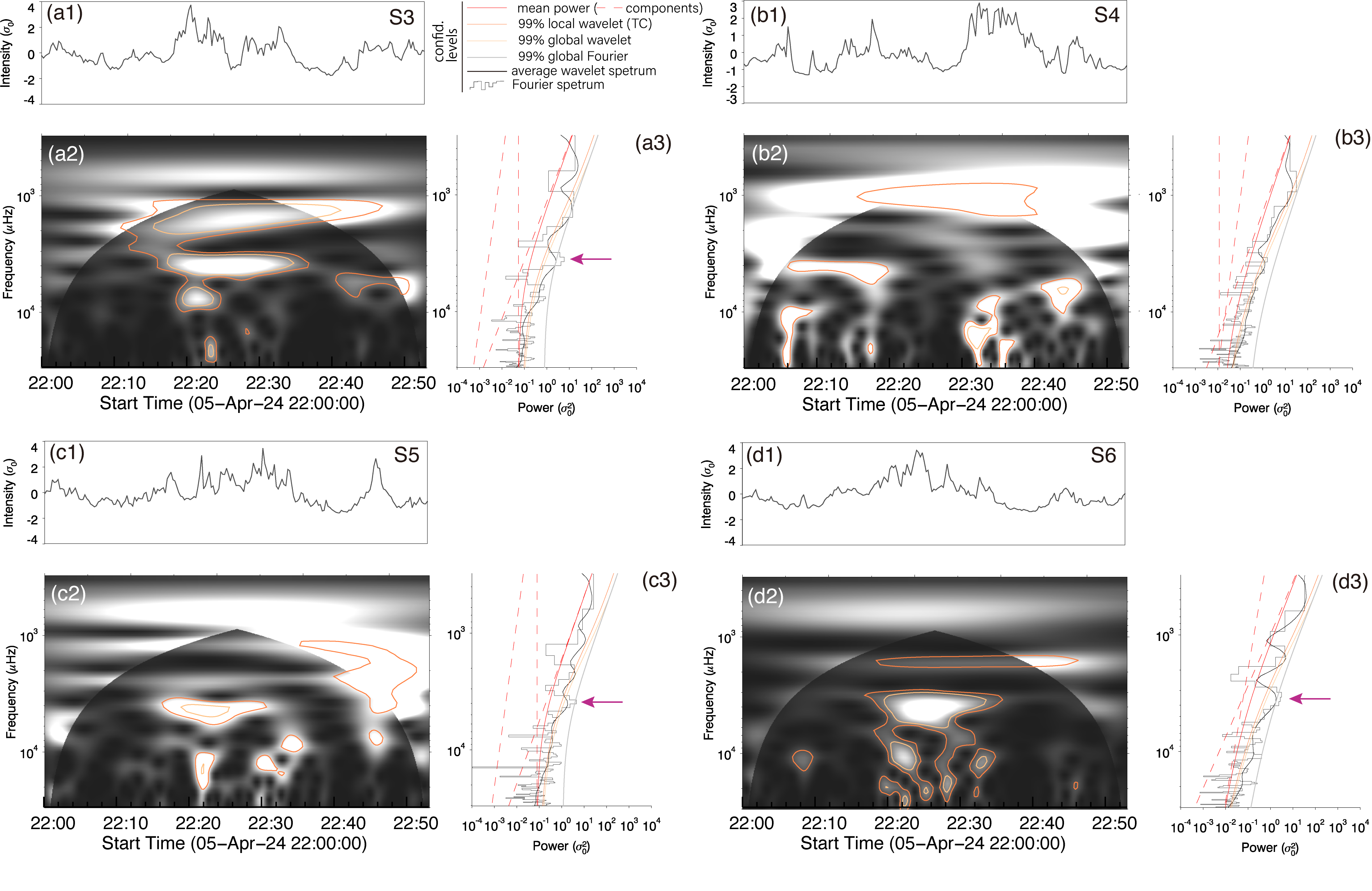}
\caption{Wavelet and Fourier analysis of the temporal evolution for plasma outflows (Fig. 3(i3)) and curtain feature (Fig. 4(i1)$-$4(i3)). The variance-normalized light curve is displayed in the upper left corner of each panel. Panels (a2), (b2), (c2), and (d2) illustrate the whitened Morlet wavelet spectrum, which was normalized according to the background power model. The Fourier power spectra (black histogram), time-averaged wavelet spectrum (thick line), and fitted noise model (red line), along with its components (dashed red line), are shown in panels (a3), (b3), (c3), and (d3). The 99\% local and global confidence levels with fitted noise are shown as dark and light orange curves, respectively. The gray line indicates the 99\% global confidence level for the Fourier spectrum. The peak of the Fourier power is labeled with pink arrows.
\label{fig:general}}
\end{figure*}
To investigate the periodic oscillation of plasma outflows along the jet spire and the curtain feature, we applied a wavelet and Fourier analysis in Fig. 7 to four light curves, following the method introduced by \citet{1998BAMS...79...61T}and \citet{2016ApJ...825..110A}. The positions from which the light curves were extracted are marked by the dashed red lines in Fig. 3(i3) and Figs. 4(i1)$-4$(i3). We chose to fit the fast Fourier transform (FFT) of each light curve using the following power-law equation:
 \begin{equation}
 \sigma(\nu)=A\nu^{s}+BK_\rho(\nu)+C,
 \end{equation}
where $\nu$ is the Fourier frequency, A, B, and C are constants, the second term accounts for the possible presence of pulses in the time series, and $K_\rho(\nu)$ is the Kappa function,
\begin{equation}
K_\rho(\nu)=\left(1 + \frac{\nu^{2}}{\kappa \rho^{2}}\right)^\frac{-\kappa+1}{2},
\end{equation}
where $\rho$ is the width of the power spectral distribution (PSD), and $\kappa$ defines the extent of its high-frequency wing. 
\par
Figs. 7(a3)$-$7(d3) show that the global 99\% Fourier confidence level is higher than the local and global 99\% wavelet confidence levels. The global 99\% wavelet confidence
levels are higher than the local 99\% wavelet confidence level. This result is consistent with findings reported in previous studies \citep[e.g.,][]{2016ApJ...825..110A,2020A&A...634A..63K,2022A&A...662A...7L,2022A&A...664A..28M}. In Figs. 7(a3), 7(c3), and 7(d3), a slope variation in the Fourier spectrum forms a peak around {$5\times10^3\,\mu\text{Hz}$}, corresponding to a period of 200 s. This indicates that periodic interchange reconnection occurred in the curtain structure (Figs. 7(b3), 7(c3), and 7(d3)), which is comparable to the periodicity of the interchange reconnection at the jet spire (Fig. 7(a3)), with both being approximately 200 s. During the eruption, the upward movement of cold plasma obscures the hot plasma outflows. This might explain the lack of a distinct periodicity in Fig. 7(b3). In this scenario, the periodicity of the interchange reconnection is maybe also underestimated.

\section{Conclusion and discussion}

Using high spatial resolution observations from the Extreme Ultraviolet Imager (EUI) on board the Solar Orbiter,  we reported the direct imaging of three current sheets (CSs) associated with recurrent interchange reconnection near the null point within a fan-spine-like system. The three current sheets occurring near the null point were modulated by emerged small magnetic structures. CS1 was generated by the interaction between emerging magnetic loops and the ambient magnetic field. An X point at the reconnection site and plasma blobs within CS1 were observed. CS2 appeared to be affected by the eruption of filament 1 on the side of the fan-spine-like topology, lasting about 6 minutes with a maximum length of 5 Mm and a width of 0.8 Mm. The formation mechanism of CS3 is similar to the flare current sheet in the breakout jet model. CS3 lasted 17 minutes, with a maximum length of 7.4 Mm and a width of 0.6 Mm. The estimated reconnection rates are approximately 0.12 to 0.2 for the CS2 and 0.12 to 0.17 for the CS3. A curtain-like feature with a width of 1.7 Mm arose during the dynamic evolution of the fan-spine system. Periodic interchange reconnection occurred in the curtain, with a periodicity comparable to that of interchange reconnection at the jet spire, both being about 200 s. Additionally, the plasma outflows from periodic reconnections, regardless of where they occurred, nearly encompassed the evolution of the fan-spine system and showed a gradual physical process similar to that of interchange reconnection in a pseudostreamer. We suggest that the dynamics of recurrent magnetic reconnection in such a small fan-spine system might be more complex than we realized before. Our results revealed a self-similarity between small-scale fan-spine structures and large-scale pseudostreamer structures.
\par
We found that during the dynamic evolution of the fan-spine system, complex magnetic field interchange reconnection occurred near the null point. Specifically, the appearance of CS1 resulted from the interchange reconnection between the newly emerging magnetic flux and background fields (Fig. 2); the transition from CS2 to CS3 was accompanied by exchanges in reconnection outflows and inflows (Fig. 3); interchange reconnection occurred between the closed magnetic loop under the fan dome and the surrounding open field, forming a curtain-like structure in the reconnection outflow region of CS3 (Fig. 4); and periodic reconnection occurred near the null point, including on the jet spire and the curtain feature. These results suggest that the magnetic reconnection in a fan-spine topology seems to be more complex than single null-point reconnection, possibly involving QSL reconnection or separators (i.e., a series of nulls). \citet{2024ApJ...962L..38D} reported that two null points can be connected by an elongated sheetlike quasi-separatrix layer (QSL) through the magnetic null-point-tracing method and the calculation of the 3D squashing factor Q. This result indicates the possible existence of multiple magnetic null points in the fan-spine system.

\par
High-resolution EUI 174 \AA~images revealed that persistent plasma outflows occur near the magnetic null point. The velocity of these plasma outflows ($\sim$ 64-90$\speed$) is somewhat lower than that of the filament 1 eruption (i.e., 186$\speed$), indicating that gentle reconnection occurs persistently near the magnetic null point. Recurrent reconnection that persistently occurs near the magnetic null point in the fan-spine configuration, resulting in plasma outflows, was also reported in previous studies \citep{2023NatCo..14.2107C,2024ApJ...962L..38D,2024ApJ...973...74K}. This process resembles the simulations by \citet{2022ApJ...935L..21N,2023ApJ...958L..38N}, where stochastic granular motions were suﬃcient to stress the small-scale fan-spine structure, triggering sustained reconnection at the coronal null point, which naturally resulted in persistent jetting activity in the form of recurrent gentle plasma outflows. Gentle reconnection or gentle plasma outflows most likely result from the interchange reconnection between magnetic flux under the fan dome and the ambient open magnetic field lines. These gradual physical processes are similar to the interchange reconnection that occurs in the pseudostreamer structure \citep[e.g.,][]{2014ApJ...787..145M} and probably plays a critical role in the continuous energy and material injection into the high corona.
\par
The observation that sparked our interest in this event was the identification of two distinct current sheets (CS2 and CS3) at disparate locations. It is pertinent to note that changes in magnetic pressure or gas pressure can lead to reconnection reversals \citep{2009A&A...494..329M}, resulting in the formation of oscillatory current sheets around the null point, as demonstrated by \citet{2019ApJ...874..146H}. Their study focused on oscillatory magnetic reconnection preceding a jet within the breakout current sheet. Here, the appearance of CS2 and CS3 in sequence also indicates the reconnection reversal. This should be attributed to the dynamics described by the breakout jet model, however, which is different from the oscillatory reconnection. In simulations conducted by \citet{2017Natur.544..452W}, shear facilitated the accumulation of free energy within the filament channels along the polarity inversion line of the fan-spine magnetic configuration. As the shear progressed, the overlying magnetic field expanded upward and formed a breakout current sheet (BCS) at the null point. In this event, the appearance of CS2 was affected by the eruption of filament 1 on the side of the fan-spine-like structure. Reconnection at the CS2 then eliminated the restraining fields, resulting in eruptions of another filament \citep[filament 2; e.g.,][]{2017Natur.544..452W,2018ApJ...852...98W,2018ApJ...854..155K,2023ApJ...953..148S,2024ApJ...968..110D,2024ApJ...962L..38D,2024MNRAS.528.1094Y}. Filament 2 then rapidly reconnected with the external open field, inducing explosive interchange reconnection within the CS3 left in its wake, thereby producing the post-flare loops. The transition from CS2 to CS3 in this event is similar to the breakout jet model \citep{2017Natur.544..452W}. In the simulation studies \citep[e.g.,][]{2021ApJ...913...64S}, the reversal or transition of these current sheets was apparently not considered. Our observations also indicate that the dynamics of large-scale pseudostreamer structures might undergo similar complex processes, which might be a possible reason for the appearance of complex structures in the solar wind. 
\par
Our calculations of the width of the current sheet might overestimate the actual width because the features we observed were essentially plasma sheets embedded within the current sheet \citep[e.g.,][]{2015SSRv..194..237L}. When we consider that the thermal halo effect has very little impact, the lengths of the current sheets are generally comparable to the lengths of the plasma sheets. The calculated reconnection rate is therefore likely overestimated and only represents an upper limit of the true reconnection rate. The overall trend reflects the evolutionary characteristics of the reconnection rate, however.

\par
In large-scale pseudostreamer structures, the separatrix curtain formed by the separatrix fan surface from the central null is orthogonal to the other two surfaces and is bounded by their respective spine field lines \citep{2011ApJ...731..111T}. Compared to the fan-spine topology (often described as a separatrix dome with a single null point), separatrix curtains involve multiple null points that are connected by separators. When we consider the three-dimensional invariance of the fan-spine structure in a two-dimensional geometry, the three-dimensional null point is replaced with a separator, and the spine line is replaced with a spine surface.  Within the three-dimensional magnetic reconnection framework, reconnection is not confined to null points, but occurs throughout a finite diffusion region \citep{2022LRSP...19....1P}. In this scenario, the fan-spine structure resembles a scaled-down version of the pseudostreamer structure \citep[e.g.,][]{2013ApJ...764...87L,2024ApJ...975..168W}. In the small-scale fan-spine topology, multiple nulls and connecting separators allow a more frequent and complex interchange reconnection between open and closed magnetic fluxes, similar to the pseudostreamer simulated by \citet{2021ApJ...913...64S}. Consistent with our previous findings \citep{2024ApJ...962L..38D}, our current observations revealed the occurrence of a complex and frequent interchange reconnection in the coronal jets between open and closed magnetic fluxes, likely mediated by multiple null points. Concurrently, the coronal jet expelled plasma with a curtain-like jet spire and a complex current sheet evolution. These results provide evidence that this coronal jet occurred in a small-scale separatrix curtain, resembling a scaled-down version of the pseudostreamer structure. Similarly, \citet{2022A&A...664A..28M} recently reported analogous phenomena, including a curtain-like jet spire and numerous complex morphological changes, in a highly dynamic small-scale event observed by EUI in a polar coronal hole. Moreover, \citet{2025A&A...702A.201L} also reconstructed a separatrix curtain-like magnetic topology in small-scale recurrent jets. These high-resolution EUI imaging observations jointly reinforce the view of the across-scale self-similarity of magnetic separatrix curtains and their associated interchange reconnection processes from the large-scale pseudostreamers down to tiny coronal jets.

\begin{acknowledgements}
    The authors are grateful for the helpful discussions with Prof. Hui Tian from Peking University and Yuandeng Shen from Harbin Institute of Technology, and the anonymous referee’s valuable comments and suggestions. This work is supported by the Strategic Priority Research Program of the Chinese Academy of Sciences,  Grant No. XDB0560000, NSFC grant 12403065 \& 12425301, Beijing Natural Science Foundation (1244053). X.L.Y was supported by the National Science Foundation of China (NSFC) under Nos. 12325303, Yunnan Key Laboratory of Solar Physics and Space Science under No. 202205AG070009, and Yunnan Fundamental Research Projects under Nos. 202301AT070347 and 202301AT070349. H.C.C. was supported by NSFC grant (12573061) and the Yunnan Fundamental Research Projects under Nos. 202401CF070165. J.C.H. was supported by the NSFC grant 12173084, the CAS “Light of West China” Program, and the Yunnan Science Foundation of China (202401AT070071). Z.Y.H. was supported by NSFC grant 12303057. J.C.W. was supported by the Yunnan Fundamental Research Projects under Nos. 202501AW070002. The authors are grateful for the data provided by the Solar Orbiter science teams. Solar Orbiter is a mission of international cooperation between ESA and NASA, operated by ESA. The EUI instrument was built by CSL, IAS, MPS, MSSL/UCL, PMOD/WRC, ROB, LCF/IO with funding from the Belgian Federal Science Policy Office (BELSPO/PRODEX PEA 4000112292); the Centre National d’Etudes Spatiales (CNES); the UK Space Agency (UKSA); the Bundesministerium für Wirtschaft und Energie (BMWi) through the Deutsches Zentrum für Luftund Raumfahrt (DLR); and the Swiss Space Office (SSO).
\end{acknowledgements}

\bibliographystyle{aa} 
\bibliography{ref}

\end{document}